\documentclass[a4paper,10.5pt]{article}
\usepackage{amssymb}
\usepackage{amsmath}
\usepackage{latexsym}
\usepackage{array}
\usepackage{bm}
\usepackage{theorem}
\usepackage{graphicx}
\usepackage{exscale}

\theoremstyle{break}
\newtheorem{Theorem}{Theorem}[section]
\newtheorem{Proposition}[Theorem]{Proposition}
\newtheorem{Lemma}[Theorem]{Lemma}

\newtheorem{Example}[Theorem]{Example}

\newtheorem{Definition}[Theorem]{Definition}

\def\Proof{\hfil\break{\bf Proof}\;\;\;\;}

\def\DIS{\displaystyle}

\def\n{{\boldsymbol{n}}}
\def\m{{\boldsymbol{m}}}

\def\e{\boldsymbol{e}}
\def\x{\boldsymbol{x}}
\def\y{\boldsymbol{y}}
\def\z{\boldsymbol{z}}

\def\bmu{\boldsymbol{\mu}}
\def\bX{\boldsymbol{X}}

\def\mF{\mathcal{F}}

\def\X{\mbox{\boldmath $X$}}

\def\F{\boldsymbol{F}}

\def\Z{\mathbb{Z}}

\def\qed{\hfill\hbox{\rule[-2pt]{3pt}{6pt}}}

\begin{document}

\title
{Toda type equations over multi-dimensional lattices} 
\author
{Ryo Kamiya$^1$, Masataka Kanki$^2$, Takafumi Mase$^1$, Naoto Okubo$^3$, Tetsuji Tokihiro$^1$\\
\small $^1$ Graduate School of Mathematical Sciences,
\small the University of Tokyo, 3-8-1 Komaba, Tokyo 153-8914, Japan\\
\small $^2$ Department of Mathematics,
\small Kansai University, 3-3-35 Yamate, Osaka 564-8680, Japan\\
\small $^3$ Department of Physics and Mathematics,
\small Aoyama Gakuin University, 5-10-1 Fuchinobe, Kanagawa 252-5258, Japan}

\date{}

\maketitle

\begin{abstract}
We introduce a class of recursions defined over the $d$-dimensional integer lattice.
The discrete equations we study are interpreted as higher dimensional extensions to the discrete Toda lattice equation.
We shall prove that the equations satisfy the coprimeness property, which is one of integrability detectors analogous to the singularity confinement test.
While
the degree of their iterates grows exponentially, their singularities exhibit a nature similar to that of integrable systems in terms of the coprimeness property.
We also prove that the equations can be expressed as mutations of a seed in the sense of the Laurent phenomenon algebra.
\end{abstract}

\section{Introduction}
The Toda lattice was invented by Morikazu Toda fifty years from now as a mathematical model of a one-dimensional chain of equal particles connected by springs with nonlinear (having  an exponential term) potential energy \cite{Toda}. He discovered the exact two-soliton interaction of the system.
The Toda lattice is undoubtedly one of the most important integrable equations in the field of mathematical physics.
One of the two-dimensional extensions of the Toda lattice was first invented by
A. V. Mikhailov in 1979 \cite{Mikhailov}, and its elliptic form was studied in \cite{Fordy-Gibbons}.
There has been a lot of attention to the Toda lattice type equations not only in the field of mathematical physics but also in the field of quantum physics and wide areas of technology until now.
Later, difference analogues of these equations have started to draw keen attention,
mainly motivated by Ryogo Hirota's works on difference schemes of integrable equations.
A difference analogue of the two-dimensional Toda lattice was discovered by Hirota and his co-workers \cite{HTI}, and the detailed study is found in \cite{Tsujimoto}.

In this article we focus on several discrete equations (difference-difference equations) closely related to the two-dimensional Toda lattice equation.
There has been a discussion on the definition of integrability of discrete equations or discrete dynamical systems, and several integrability detectors have been invented.
The singularity confinement test \cite{SC} was first proposed as an analogue of the Painlev\'{e} test for differential equations.
A discrete equation is said to pass the singularity confinement test (SC test), if
the singularities of the equations are resolved after a finite number of iterations:
i.e., the information on the initial variables are recovered.
The SC test has been successfully used to detect the integrability of many discrete equations and to discover several discrete counterparts of the Painlev\'{e} equations \cite{dP}.
However, it has later been found that the SC test has counter-examples. For example, the Hietarinta-Viallet equation passes the SC test whilst it is non-integrable in the sense that it has chaotic orbits of iterations \cite{HV}.
Another famous integrability criterion is the zero algebraic entropy test, which asserts that a discrete equation is integrable if and only if its algebraic entropy is zero \cite{BV}. The algebraic entropy is a non-negative real number associated with discrete equations. When the degree of the iterations of an equation has polynomial growth, the algebraic entropy is zero, while, when the degree grows exponentially, the algebraic entropy is positive.
It is now believed to be one of the most reliable integrability tests, and therefore,
we hire the zero algebraic entropy (or equivalently the polynomial degree growth) as the `definition' of integrability in this article.

Recently there have been several attempts to further investigate 
discrete systems through the singularity analysis originating from the SC test.
One of them is the full-deautonomisation scheme, in which the behaviour of the nonautonomous coefficients of an equation is correlated to the induced linear action on the Picard group of the space of initial conditions
and therefore to the algebraic entropy
\cite{Redemption,Mase,redeeming}.
This technique was also applied to a nonintegrable lattice equation with confined singularities \cite{WMRG}.
The second is the coprimeness property proposed by some of the authors.
A discrete equation is said to have the coprimeness property, if arbitrary two iterates of the equation are mutually coprime when they are separated enough.
By studying the property for the discrete KdV equation and the Somos sequences, it has been found that
the satisfaction of the coprimeness property, together with the Laurent property and the irreducibility property, is quite similar to passing the SC test \cite{dKdVSC2}.
In fact, the coprimeness property is considered to be an algebraic re-interpretation (and in some cases the refinement) of the SC test.
In previous works, we have studied coprimeness-preserving extensions to the discrete
Toda lattice equation and the two-dimensional discrete Toda lattice equation \cite{2dtoda1}:
\begin{equation}
\tau_{t+1,n,m+1}\tau_{t-1,n+1,m}=\tau_{t,n+1,m}^{k_1}\tau_{t,n,m+1}^{k_2}+\tau_{t,n,m}^{l_1}\tau_{t,n+1,m+1}^{l_2}\qquad (k_1, l_1,k_2,l_2 \in \Z_+).
\label{pDToda_polinear_eq}
\end{equation}
Note that if we set the indices $k_1,k_2,l_1,l_2$ equal to $1$, we recover the two-dimensional discrete Toda lattice.
In \cite{2dtoda1}, some of the authors have proved the Laurent property, the irreducibility and the coprimeness of the equations.
The previous results are briefly reviewed in Appendix \ref{2dfactorappendix}.
In this manuscript, we study further extensions to the coprimeness-preserving two-dimensional Toda lattice equation:
\begin{equation}
\tau_{t+1,\n}\tau_{t-1,\n}=\prod_{i=1}^a\tau_{t,\n+\e_i}^{k_i}\tau_{t,\n-\e_i}^{l_i}+\prod_{i=a+1}^{a+b}\tau_{t,\n+\e_i}^{k_i}\tau_{t,\n-\e_i}^{l_i}\qquad (k_i, l_i \in \Z_+),
\label{pNDToda_polinear_eq}
\end{equation}
where, $a,b$ are non-negative integers, and $\n=(n_1,n_2,...,n_a,n_{a+1},...,n_{a+b})\in \Z^{a+b}$ is an $(a+b)$-dimensional integer lattice point, and $\e_i$ ($i=1,2,...,a+b$) are unit vectors in $\Z^{a+b}$ from $\e_1=(1,0,\cdots,0)$ to $\e_{a+b}=(0,\cdots,0,1)$.
The equation \eqref{pNDToda_polinear_eq} is considered to be a time one-dimensional,
space $(a+b)$-dimensional discrete equation.
When $a=b=1$,  equation \eqref{pNDToda_polinear_eq} is equivalent to the
coprimeness-preserving extensions to the two-dimensional Toda lattice equation \eqref{pDToda_polinear_eq}.
The degree growth of the iterates $\tau_{t,n,m}$ of \eqref{pDToda_polinear_eq} is proved to be exponential unless $k_1=k_2=l_1=l_2=1$ in our previous work \cite{2dtoda1,2dTodaFactorize}.
Similarly, it is easy to prove that the degrees deg$\, \tau_{t,\n}$ of \eqref{pNDToda_polinear_eq}
grow exponentially with respect to $t$, unless $(a,b)=(1,1)$ and $k_1=k_2=l_1=l_2=1$.
Therefore the Toda type equation \eqref{pNDToda_polinear_eq} is non-integrable in the sense of algebraic entropy test.
Nevertheless the equations of the type \eqref{pDToda_polinear_eq} and \eqref{pNDToda_polinear_eq} have several properties which
are closely related to the singularities: i.e.,
the Laurent property and the coprimeness property, and an LP algebraic expression.
Our aim in this paper is to investigate the properties of \eqref{pNDToda_polinear_eq},
in terms of its coprimeness property and its expression by the Laurent Phenomenon (LP) algebra\cite{LP}, which is one type of generalization of the cluster algebra\cite{FZ}.
These results indicate that although the equation \eqref{pNDToda_polinear_eq} is non-integrable in the sense of degree growth, it has singularity structures similar to those of integrable systems.
This paper is organized as follows: In section \ref{section2}, let us prove that
$\tau_{t,\n}$ is a Laurent polynomial of the initial variables with integer coefficients.
Moreover, we shall prove that $\tau_{t,\n}$ is irreducible in the ring of Laurent polynomials, and that two distinct iterates are coprime, if the right hand side is not decomposable.
In section \ref{section3}, we shall present an LP algebraic interpretation of equation \eqref{pNDToda_polinear_eq}.

\section{Coprimeness of higher dimensional Toda lattice} \label{section2}
Let us take the initial variables as $\tau_{0,\n}$ and $\tau_{1,\n}$ with $\n\in\mathbb{Z}^{a+b}$ and consider the evolution of the equation towards $t\ge 2$.
To simplify notation, let us define $x_{\n}:=\tau_{0,\n}$, $y_{\n}:=\tau_{1,\n}$,
$z_{\n}:=\tau_{2,\n}$, $u_{\n}:=\tau_{3,\n}$, $v_{\n}:=\tau_{4,\n}$.
We also define the sets of iterates for a certain $t$: e.g., $\x:=\{x_{\n}\}_{\n\in\Z^{a+b}}$, $\y:=\{y_{\n}\}_{\n\in\Z^{a+b}}$, and so on.
We use the notation $\x^{\pm}$ (resp. $\y^{\pm}$) to denote the set of variables and their inverse elements $\{x_{\n}, x_{\n}^{-1}\}_{\n\in\Z^{a+b}}$ (resp. $\{y_{\n}, y_{\n}^{-1}\}_{\n\in\Z^{a+b}}$).
Let us state our main theorem:
\begin{Theorem} \label{mainthm}
Suppose that
\begin{equation}
\operatorname{GCD} \left\{ k_i, l_i \mid i = 1, \ldots, a + b \right\} =2^r, \label{irredcond1}
\end{equation}
where $r$ is a non-negative integer, and let us define the ring of Laurent polynomials of the initial variables $\x\cup\y$ as $R:=\mathbb{Z}[\x^{\pm}, \y^{\pm}]$.
Then, for every $t\ge 0$ and $\n\in\mathbb{Z}^{a+b}$,
the iterate $\tau_{t,\n}$ of the equation \eqref{pNDToda_polinear_eq} belongs to $R$ and is irreducible in $R$.
Moreover, the two distinct iterates $\tau_{t,\n}$ and $\tau_{t',\n'}$ with $(t,\n)\neq (t',\n')$
are always coprime in $R$.
\end{Theorem}
The condition \eqref{irredcond1} is equivalent to
the irreducibility of
\begin{equation}\label{irred_assumption}
\prod_{i=1}^a X^{k_i} Y^{l_i}+\prod_{i=a+1}^{a+b} X^{k_i} Y^{l_i}
\end{equation}
in $\mathbb{Z}[X,Y]$.
Let us prepare two propositions to prove theorem \ref{mainthm}.
\begin{Proposition}\label{trans_prop}
Let us suppose that an iterate $\tau_{t,\n}$ is a Laurent polynomial in $R$, and that $\tau_{t,\n}$ is irreducible in $R$.
Then every iterate $\tau_{t,\n'}$ with $\n' \ne \n$ is also irreducible and is coprime with $\tau_{t,\n}$． 
\end{Proposition}
\Proof
From the translational symmetry of the iterates，we immediately obtain the irreducibility of $\tau_{t,\n'}$. The coprimeness is proved by reductio ad absurdum.
If the two iterates $\tau_{t,\n'}$ and $\tau_{t,\n}$ are not coprime with each other,
from the irreducibility of both $\tau_{t,\n}$ and $\tau_{t,\n'}$,
we have $\tau_{t,\n'}=c_{\n,\n'}\tau_{t,\n}$ where $c_{\n,\n'}$ is a unit element in $R$. Note that $c_{\n,\n'}$ is divisible by a certain $y_{\boldsymbol{m}}$,
which does not appear in $\tau_{t,\n'}$.
Now let us write $\n-\n'$ as $\n-\n'=\sum_{j=1}^{a+b} k_j\e_j$ and pick up one $j$ such that $|k_j| \ge |k_i|$ for all $i \in \{1,2,...,a+b\}$. For simplicity we reassign the subscripts
so that $j=1$, and assume without loss of generality that $k_1>0$.
From here on in this proof, let us substitute $y_{\n+(t-1)\e_1}=0$ and substitute $1$ for all the other initial variables except for $y_{\n+(t-1)\e_1}$.
Then $c_{\n,\n'}$, which is a monomial, must become either $0$ or $\infty$.
Since $\tau_{t,\n'}$ does not depend on $y_{\n+(t-1)\e_1}$ as a rational function of the initial variables,
the term $c_t :=\tau_{t,\n}|_{\x\cup \y=\{1\}}$ satisfies the following recurrence relation when $1$ is substituted for the initial variables:
\begin{equation}\label{c_eq}
c_{t+1}=\frac{1}{c_{t-1}}\left(c_t^{\sum_{i=1}^a k_i+l_i}+c_t^{\sum_{j=a+1}^{a+b} k_j+l_j}     \right),
\end{equation}
with $c_0=1$, $c_1=1$, $c_2=2$.
The iterate $\tau_{t,\n}$ satisfies the same recurrence as \eqref{c_eq} with $c_1=1$, $c_2=1$.
Therefore $\tau_{t,\n},\,\tau_{t,\n'}>0$.
This is a contradiction with the fact that $\tau_{t,\n'}=c_{\n,\n'}\tau_{t,\n}$ and that $c_{\n,\n'}=0$ or $\infty$.
\qed
\begin{Proposition}\label{order_prop}
If the two iterates $\tau_{t,\n},\, \tau_{t',\n'} \in R$ with $t \ne t'$ are both irreducible in $R$, they are mutually coprime in $R$.
\end{Proposition}
\Proof
Let us take the initial values as $x_{\n}=y_{\n}=1$ for all $\n\in\Z^{a+b}$, and define
$c_t:=\tau_{t,\n}$ for these initial values. Then $c_t$ satisfies \eqref{c_eq} with $c_0=c_1=1$. Therefore $c_t$ is strictly increasing with respect to $t$ for $t\ge 2$.
However, if $\tau_{t,\n}$ and $\tau_{t',\n'}$ are not mutually coprime, these two iterates must be equivalent except for a monomial multiple. Therefore $c_t=c_{t'}$, which contradicts the strictly increasing property.
\qed

\noindent
\textbf{Proof of theorem \ref{mainthm}}
Proof is done by induction with respect to $t$.
\begin{enumerate}
\item The case of $t=2$:
If $\n \ne \n'$, the two iterates $z_{\n}$ and $z_{\n'}$ are both non-monomial irreducible
Laurent polynomials in $R$ and are mutually coprime.
\par
$\because$ )
Since
\[
z_{\n}=\frac{1}{x_{\n}}\left(\prod_{i=1}^ay_{t,\n+\e_i}^{k_i}y_{t,\n-\e_i}^{l_i}+\prod_{i=a+1}^{a+b}y_{t,\n+\e_i}^{k_i}y_{t,\n-\e_i}^{l_i}\right),
\]
we have that $z_{\n}$, $z_{\n'}$ are both irreducible using the irreducibility of
\eqref{irred_assumption}.
Also $z_{\n}$ is not a monomial and thus is not a unit.
On the other hand, there exists at least one term that is in $z_{\n}$ but not in $z_{\n'}$. 
Thus $z_{\n'}$ cannot be equivalent to $z_{\n}$ except for a Laurent monomial multiple.
Therefore these two iterates are mutually coprime.
\qed

\item The case of $t=3$:
If $\n \ne \n'$, the two iterates $u_{\n}$ and $u_{\n'}$ are mutually coprime irreducible Laurent polynomials in $R$.
Moreover，the iterate $u_{\n}$ is coprime with $z_{\n'}$ for all $\n'$.
\par
$\because$ ) We have
\begin{equation}
u_{\n}=\frac{1}{y_{\n}}\left(\prod_{i=1}^az_{\n+\e_i}^{k_i}z_{\n-\e_i}^{l_i}+\prod_{i=a+1}^{a+b}z_{t,\n+\e_i}^{k_i}z_{\n-\e_i}^{l_i}\right). \label{uneq1}
\end{equation}
From lemma \ref{lem2previous} on the variable transformation of Laurent polynomials that has first appeared in \cite{dKdVSC2}, we have the following factorization of $u_{\n}$:
\[
u_{\n}=\left( \prod_{\m\in\mathbb{Z}^2} z_{\m}^{\alpha_{\m}}\right) f_{irr},
\]
where each $\alpha_{\m}$ is a non-negative integer, and $f_{irr}$ is an irreducible Laurent polynomial in $R$. Note that the product is essentially a finite product.
All $z_{\n}$'s are mutually coprime irreducible Laurent polynomials from the induction hypothesis and are not monomials. Thus we have  $\alpha_{\m}=0$ for all $\m$ with $\m = \n \pm \e_i$ ($i=1,2,...,a+b$) from \eqref{uneq1}.
Next we shall prove that $\alpha_{\m}=0$ for every $\m\neq \n\pm \e_i$ $(i=1,2,\cdots ,a+b)$.
For $\m$ with $\m\neq \n\pm \e_i$, the iterate $z_{\m}$ depends on $y_{\m\pm\e_i}$,
but $u_{\n}$ does not depend on $y_{\m\pm\e_i}$.
The iterate $z_{\n_k}$ is composed of a sum of two terms, which are products of some iterates from $\y$, and therefore is not a unit.
Therefore $z_{\m}$ and $u_{\n}$ are coprime and we have $\alpha_{\m}=0$．
Thus $u_{\n}$ is irreducible.
From proposition \ref{order_prop}, the iterate $u_{\n}$ is coprime with $z_{\n'}$ for all $\n'$.
Also from proposition \ref{trans_prop}, every pair from the set $\boldsymbol{u}$ is mutually coprime.
\qed
\item The case of $t=4$ (part 1):
We prove $v_{\n}\in R$. First we compute
\begin{align*}
v_{\n}z_{\n}&=\prod_{i=1}^a \left\{\frac{1}{y_{\n+\e_i}}\left(\prod_{i'=1}^az_{\n+\e_i+\e_{i'}}^{k_{i'}}z_{\n+\e_i-\e_{i'}}^{l_{i'}}+\prod_{j'=a+1}^{a+b} z_{\n+\e_i+\e_{j'}}^{k_{j'}}z_{\n+\e_i-\e_{j'}}^{l_{j'}}   \right)\right\}^{k_i}\\
&\quad \times \left\{\frac{1}{y_{\n-\e_i}}\left(\prod_{i'=1}^az_{\n-\e_i+\e_{i'}}^{k_{i'}}z_{\n-\e_i-\e_{i'}}^{l_{i'}}+\prod_{j'=a+1}^{a+b} z_{\n-\e_i+\e_{j'}}^{k_{j'}}z_{\n-\e_i-\e_{j'}}^{l_{j'}}   \right)\right\}^{l_i}\\
&+\prod_{j=a+1}^{a+b} \left\{\frac{1}{y_{\n+\e_j}}\left(\prod_{i'=1}^az_{\n+\e_j+\e_{i'}}^{k_{i'}}z_{\n+\e_j-\e_{i'}}^{l_{i'}}+\prod_{j'=a+1}^{a+b} z_{\n+\e_j+\e_{j'}}^{k_{j'}}z_{\n+\e_j-\e_{j'}}^{l_{j'}}   \right)\right\}^{k_j}\\
&\quad \times \left\{\frac{1}{y_{\n-\e_j}}\left(\prod_{i'=1}^az_{\n-\e_j+\e_{i'}}^{k_{i'}}z_{\n-\e_j-\e_{i'}}^{l_{i'}}+\prod_{j'=a+1}^{a+b} z_{\n-\e_j+\e_{j'}}^{k_{j'}}z_{\n-\e_j-\e_{j'}}^{l_{j'}}   \right)\right\}^{l_j}.
\end{align*}
By continuing the calculation of the right hand side modulo $z_{\n}$ we obtain
\begin{align*}
v_{\n}z_{\n}&\equiv \prod_{i=1}^a \left\{\frac{1}{y_{\n+\e_i}}\left(\prod_{j'=a+1}^{a+b} z_{\n+\e_i+\e_{j'}}^{k_{j'}}z_{\n+\e_i-\e_{j'}}^{l_{j'}}   \right)\right\}^{k_i}
\left\{\frac{1}{y_{\n-\e_i}}\left(\prod_{j'=a+1}^{a+b} z_{\n-\e_i+\e_{j'}}^{k_{j'}}z_{\n-\e_i-\e_{j'}}^{l_{j'}}   \right)\right\}^{l_i}\\
&+\prod_{j=a+1}^{a+b} \left\{\frac{1}{y_{\n+\e_j}}\left(\prod_{i'=1}^az_{\n+\e_j+\e_{i'}}^{k_{i'}}z_{\n+\e_j-\e_{i'}}^{l_{i'}} \right)\right\}^{k_j}
\left\{\frac{1}{y_{\n-\e_j}}\left(\prod_{i'=1}^az_{\n-\e_j+\e_{i'}}^{k_{i'}}z_{\n-\e_j-\e_{i'}}^{l_{i'}}\right)\right\}^{l_j}\\
&=\frac{\prod_{i=1}^ay_{\n+\e_i}^{k_i}y_{\n-\e_i}^{l_i}+\prod_{j=a+1}^{a+b}y_{\n+\e_j}^{k_j}y_{\n-\e_j}^{l_j}}{\prod_{r=1}^{a+b}y_{\n+\e_r}^{k_r}y_{\n-\e_r}^{l_r} }
\left(\prod_{i=1}^a\prod_{j=a+1}^{a+b}z_{\n+\e_i+\e_j}^{k_ik_j}z_{\n+\e_i-\e_j}^{k_il_j}
z_{\n-\e_i+\e_j}^{l_ik_j}z_{\n-\e_i-\e_j}^{l_il_j}  \right)\\
&=\frac{z_{\n}x_{\n}}{\prod_{r=1}^{a+b}y_{\n+\e_r}^{k_r}y_{\n-\e_r}^{l_r} }
\left(\prod_{i=1}^a\prod_{j=a+1}^{a+b}z_{\n+\e_i+\e_j}^{k_ik_j}z_{\n+\e_i-\e_j}^{k_il_j}
z_{\n-\e_i+\e_j}^{l_ik_j}z_{\n-\e_i-\e_j}^{l_il_j}  \right).
\end{align*}
Therefore we have
\[
v_{\n}z_{\n}=z_{\n} P,
\]
where $P\in R$.
Eliminating $z_{\n}$ from both hands and using an induction hypothesis that
$z_{\n}$ is always in $R$, we have $v_{\n}\in R$.
\qed
\item The case of $t\ge 5$ (part 1):
By the same calculation as the previous case, we have
\begin{align*}
\tau_{t,\n}\tau_{t-2,\n} &\equiv
\frac{\tau_{t-2,\n} \tau_{t-4,\n}}
{\prod_{r=1}^{a+b} \tau_{t-3,\n+\e_r}^{k_r}\tau_{t-3,\n-\e_r}^{l_r} }\\
&\times
\left(\prod_{i=1}^a\prod_{j=a+1}^{a+b} \tau_{t-2,\n+\e_i+\e_j}^{k_ik_j} \tau_{t-2,\n+\e_i-\e_j}^{k_il_j}
\tau_{t-2,\n-\e_i+\e_j}^{l_ik_j} \tau_{t-2,\n-\e_i-\e_j}^{l_il_j}  \right) \mod \tau_{t-2,\n}.
\end{align*}
From the relation above, we can prove inductively that $\tau_{t,\n}\in R$, on condition that every pair from the three iterates $\tau_{t-2,\m},\,\tau_{t-3,\m'},\,\tau_{t-4,\m''}$ are coprime
for arbitrary $\m, \m',\m''\in \Z^{a+b}$.
Therefore our proof proceeds as follows: let us assume that the Laurent property and the irreducibility of $\tau_{s,\n}$ is proved for every $s\le t-1$. Then from this paragraph we conclude that $\tau_{t,\n}$ is in $R$. The rest of our tasks is to prove the irreducibility of $\tau_{t,\n}$.
When the irreducibility is proved, the mutual coprimeness of two distinct iterates is readily obtained from propositions \ref{trans_prop} and \ref{order_prop}.
\item The case of $t=4$ (part 2): The iterate $v_{\n}$ is irreducible in $R$.

Let us suppose otherwise. Then $v_{\n}$ is factored as
\[
v_{\n}=\left( \prod_{\m}z_{\m}^{\alpha_{\m}}\right) f_{irr},
\]
from lemma \ref{lem2previous}.
For a vector
$\n=(n_1,n_2,...,n_a,n_{a+1},...,n_{a+b})$, let us define
\[
m:=n_1+n_2+\cdots +n_a,\;\; n:=n_{a+1}+\cdots +n_{a+b},
\]
and let us choose an initial condition such that $\x,\,\y$ depend only on $m,n$, but do not depend on each $n_i$ $(i=1,2,\cdots,a+b)$.
We consider the reduction of \eqref{pNDToda_polinear_eq} with the restriction to $\tau_{t,\n}$ such that $\tau_{t,\n}:=\tau_{t,m,n}$ depends only on $m,n$.
To be more precise, we assume that
$\tau_{t,n_1,\cdots, n_{a+b}}=\tau_{t,n'_1,\cdots, n'_{a+b}}$ if and only if
\[
\sum_{i=1}^a n_i=\sum_{i=1}^a n'_i \ \mbox{and}\ \sum_{j=a+1}^{a+b} n_i=\sum_{j=a+1}^{a+b} n'_i.
\]

If we define the new parameters
\[
K_1:=\sum_{i=1}^ak_i,\;\;L_1:=\sum_{i=1}^al_i,\quad K_2=\sum_{j=a+1}^{a+b}k_j,\;\;
L_2:=\sum_{j=a+1}^{a+b}l_j,
\]
we have the following lower-dimensional equation:
\begin{equation} \label{cp2dtoda}
\tau_{t+1,m,n}\tau_{t-1,m,n}=\tau_{t,m+1,n}^{K_1}\tau_{t,m-1,n}^{L_1}+\tau_{t,m,n+1}^{K_2}\tau_{t,m,n-1}^{L_2},
\end{equation}
which is equivalent to the coprimeness-preserving two-dimensional discrete Toda equation \eqref{pDToda_polinear_eq}.
The coprimeness property of equation \eqref{cp2dtoda} is already proved in \cite{2dTodaFactorize}. The statement is recast in the Appendix \ref{2dfactorappendix} as theorem \ref{factorizethm}. The proof is found in \cite{2dtoda1, 2dTodaFactorize}.
Therefore $v_{\n}=v_{m,n}$ is coprime with every $z_{m',n'}$, for the initial condition
such that $\x,\,\y$ depend only on $m,n$.
Since $z_{\m}=z_{m,n}$ is not a unit, we have $\alpha_{\m}=0$.
Thus $v_{\n}$ is irreducible.

\item The case of $t\ge 5$ (part 2):
All the iterates $\tau_{t,\n} \in R$ are irreducible and mutually coprime.
Since we have the following factorization from lemma \ref{lem2previous}:
\[
\tau_{t,\n}=\left( \prod_{\m}z_{\m}^{\alpha_{\m}}\right) f_{irr},
\]
we can prove the irreducibility of $\tau_{t,\n}$ in the same manner as in the case of
$t=4$.
Note that we have used the  theorem \ref{factorizethm}
for coprimenesss-preserving two-dimensional discrete Toda equation.
The coprimeness is readily obtained from the irreducibility using  propositions
\ref{trans_prop} and \ref{order_prop}.
\end{enumerate}
Now the proof of theorem \ref{mainthm} is complete.
\qed


%
%
\section{Realization as LP algebraic object} \label{section3}
In this section, we consider an LP algebra with infinite number of cluster variables, a mutation of which is expressed by  \eqref{pNDToda_polinear_eq}.  
A brief review of LP algebras is given in Appendix~\ref{LP_Basic_facts}, and we use the terminology in it.

Let us consider cluster variables $\{\tau_{t,\n}\}$ 
($ t \in \Z_{\ge 0},\, \n \in \Z_{a+b}$) and exchange polynomials $\{X_{t,\n}\}$
\begin{equation}\label{defX}
X_{t,\n}
:=\prod_{i=1}^a\tau_{t,\n+\e_i}^{k_i}\tau_{t,\n-\e_i}^{l_i}+\prod_{i=a+1}^{a+b}\tau_{t,\n+\e_i}^{k_i}\tau_{t,\n-\e_i}^{l_i},
\end{equation}
where $\e_i$, $k_i$, $l_i$ are the same as in \eqref{pNDToda_polinear_eq}.
For the initial seed $(\x_0, \F_0)$, we take
\[
\x_0=\left(\{\tau_{0,\n}\}_{\n \in \Z^{a+b}},\ \{\tau_{1,\n}\}_{\n \in \Z^{a+b}}\right),\quad
\F_0=\left(\{X_{1,\n}\}_{\n \in \Z^{a+b}},\ \{X_{0,\n}\}_{\n \in \Z^{a+b}}\right),
\]
that is, an element of the initial seed is $(\tau_{0,\n}, X_{1,\n})$ or $(\tau_{1,\n}, X_{0,\n})$.
We wish to prove that, by mutating once at a time at all the iterates $\tau_{0,\n}$ ($\n \in \Z_{a+b}$), we have
\begin{equation}\label{LP_onestep}
(\tau_{0,\n}, X_{1,\n}) \rightarrow (\tau_{2,\n}, X_{1,\n}),\quad
(\tau_{1,\n}, X_{0,\n}) \rightarrow (\tau_{1,\n}, X_{2,\n})
\end{equation}
with $\tau_{2,\n}$ given by \eqref{pNDToda_polinear_eq}.
Repeating the same mutation at $\{\tau_{1,\n}\},\,\{\tau_{2,\n}\},\, ...$, we have a sequence of seeds as
\[
\begin{pmatrix}
(\tau_{0,\n}, X_{1,\n})\\
(\tau_{1,\n}, X_{0,\n})
\end{pmatrix}
\rightarrow
\begin{pmatrix}
(\tau_{2,\n}, X_{1,\n})\\
(\tau_{1,\n}, X_{2,\n})
\end{pmatrix}
\rightarrow
\begin{pmatrix}
(\tau_{2,\n}, X_{3,\n})\\
(\tau_{3,\n}, X_{2,\n})
\end{pmatrix}
\rightarrow
\begin{pmatrix}
(\tau_{4,\n}, X_{3,\n})\\
(\tau_{3,\n}, X_{4,\n})
\end{pmatrix}
\rightarrow
\cdots,
\]
where $\tau_{t,\n}$ are defined by \eqref{pNDToda_polinear_eq}.
Hence we find that, if \eqref{LP_onestep} holds, the extended discrete Toda lattice equation \eqref{pNDToda_polinear_eq} has the LP algebraic structure and its iterates naturally show the Laurent property. 

To avoid confusion by subscripts, we put
\[
x_\n:=\tau_{0,\n},\ y_\n:=\tau_{1,\n},\ z_\n:=\tau_{2,\n}.
\]
We also define
\[\x:=\{x_\n\}_{\n \in \Z^{a+b}},\,\y:=\{y_\n\}_{\n \in \Z^{a+b}},\,\z:=\{z_\n\}_{\n \in \Z^{a+b}}
,\quad \bX_t:=\{X_{t,\n}\}_{\n \in \Z^{a+b}}.
\]
Let us consider the mutation of the initial seed $((\x,\X_1),(\y,\X_0))$ at $x_\n$.
Since the exchange polynomial $X_{1,\n}$ does not contain $x_\n$, we have $\hat{X}_{1,\n}=X_{1,\n}$ and 
\[
\mu_{x_\n}(x_\n)=\frac{X_{1,\n}}{x_{\n}}=z_\n,
\]
where $\mu_{x_\n}$ expresses the mutation at $x_{\n}$.
For arbitrary $\m \in \Z^{a+b}$, $X_{1,\m}$ does not contain $\x_n$ and we find
\[
\mu_{x_\n}(x_\m)=x_\m\, (\m \ne \n),\quad \mu_{x_\n}(X_{1,\m})=X_{1,\m}.
\]
Now we consider the change in $X_{0,\m}$. 
If ${}^\exists i,\,\n=\m\pm \e_i$, it contains $x_\n$.
In the case $\n=\m+\e_1$,
\[
G_\m:=X_{0,\m}\Big|_{x_{\m+\e_1} \leftarrow \frac{X_{1,\m+\e_1}|_{y_{\m}=0}}{z_{\m+\e_1}}},
\]
where
\[
X_{1,\m+\e_1}\big|_{y_{\m}=0}=\prod_{i=a+1}^{a+b}y_{\m+\e_1+\e_i}^{k_i}y_{\m+\e_1-\e_i}^{l_i}
\]
and 
\[
G_\m=\left( \frac{\prod_{i=a+1}^{a+b}y_{\m+\e_1+\e_i}^{k_i}y_{\m+\e_1-\e_i}^{l_i}}{z_{\m+\e_1}}   \right)^{k_1}x_{\m-\e_1}^{l_1}\prod_{i=2}^{a}x_{\m+\e_i}^{k_i}x_{\m-\e_i}^{l_i}+\prod_{i=a+1}^{a+b}x_{\m+\e_i}^{k_i}x_{\m-\e_i}^{l_i}.
\]
Clearly $G_\m$ does not have a common factor with $X_{1,\m+\e_1}\big|_{y_{\m}=0}$. Multiplying a unit $M=z_{\m+\e_1}^{k_1}$, 
we obtain
\begin{align*}
&\mu_{x_{\m+\e_1}}\left( X_{0,\m} \right)\\
&=\left(\prod_{i=a+1}^{a+b}y_{\m+\e_1+\e_i}^{k_i}y_{\m+\e_1-\e_i}^{l_i} \right)^{k_1}x_{\m-\e_1}^{l_1}\prod_{i=2}^{a}x_{\m+\e_i}^{k_i}x_{\m-\e_i}^{l_i}+z_{\m+\e_1}^{k_1}\prod_{i=a+1}^{a+b}x_{\m+\e_i}^{k_i}x_{\m-\e_i}^{l_i}. 
\end{align*}
When we perform the second mutation at $x_{\n'}$, it is clear that
\[
\mu_{x_{\n'}}(z_\n)=z_\n,\quad \mu_{x_{\n'}}(x_\m)=x_\m\, (\m \ne \n, \n'),\quad \mu_{x_{\n'}}(X_{1,\m})=X_{1,\m}.
\]
For $\mu_{x_{\n'}}(X_{0,\m})$，there are three cases where an exchange polynomial $X_{0,\m}$ changes its form:
(a) $\n'=\m-\e_1$;
(b) $\n'=\m\pm \e_i$ ($i=2,3,...,a$);
(c) $\n'=\m\pm \e_i$ ($i=a+1,...,a+b$).
For the cases (a), (b), no common factor appears when we construct new exchange polynomials by the substitution $x_{\n'} \leftarrow \frac{X_{1,\n'}|_{y_\m=0}}{z_{\n'}}$.
For example, 
\begin{align*}
&\mu_{x_{\m-\e_1}}\left( \mu_{x_{\m+\e_1}}\left( X_{0,\m} \right)\right) \\
&=\left(\prod_{i=a+1}^{a+b}y_{\m+\e_1+\e_i}^{k_i}y_{\m+\e_1-\e_i}^{l_i} \right)^{k_1}
\left(\prod_{i=a+1}^{a+b}y_{\m-\e_1+\e_i}^{k_i}y_{\m-\e_1-\e_i}^{l_i} \right)^{l_1}\prod_{i=2}^{a}x_{\m+\e_i}^{k_i}x_{\m-\e_i}^{l_i}\\
&\quad +z_{\m+\e_1}^{k_1}z_{\m-\e_1}^{l_1}\prod_{i=a+1}^{a+b}x_{\m+\e_i}^{k_i}x_{\m-\e_i}^{l_i} \\
&\mu_{x_{\m+\e_2}}\left( \mu_{x_{\m+\e_1}}\left( X_{0,\m} \right)\right) \\
&=\left(\prod_{i=a+1}^{a+b}y_{\m+\e_1+\e_i}^{k_i}y_{\m+\e_1-\e_i}^{l_i} \right)^{k_1}
\left(\prod_{i=a+1}^{a+b}y_{\m+\e_2+\e_i}^{k_i}y_{\m+\e_2-\e_i}^{l_i} \right)^{k_2}x_{\m-\e_1}^{l_1}x_{\m-\e_2}^{l_2}\prod_{i=3}^{a}x_{\m+\e_i}^{k_i}x_{\m-\e_i}^{l_i}\\
&\quad +z_{\m+\e_1}^{k_1}z_{\m+\e_2}^{k_2}\prod_{i=a+1}^{a+b}x_{\m+\e_i}^{k_i}x_{\m-\e_i}^{l_i} 
\end{align*}
On the other hand, in the case (c), if we consider $\n'=\m+\e_{a+1}$ as an example,
\begin{align*}
&\mu_{x_{\m+\e_1}}
\left( X_{0,\m} \right)\big|_{x_{\n'} \leftarrow \frac{X_{1,\n'}|_{y_{\m}=0}}{z_{\n'}}}\\
&=\left(\prod_{i=a+1}^{a+b}y_{\m+\e_1+\e_i}^{k_i}y_{\m+\e_1-\e_i}^{l_i} \right)^{k_1}
x_{\m-\e_1}^{l_1}\prod_{i=2}^{a}x_{\m+\e_i}^{k_i}x_{\m-\e_i}^{l_i}\\
&\quad +\left(\frac{\prod_{i=1}^{a}y_{\m+\e_{a+1}+\e_i}^{k_i}y_{\m+\e_{a+1}-\e_i}^{l_i}}{z_{\m+\e_{a+1}}} \right)^{k_{a+1}}z_{\m+\e_1}^{k_1}x_{\m-\e_{a+1}}^{l_{a+1}}\prod_{i=a+2}^{a+b}x_{\m+\e_i}^{k_i}x_{\m-\e_i}^{l_i}. 
\end{align*}
Hence there exists a common factor
$y_{\m+\e_1+\e_{a+1}}^{k_1k_{a+1}}$.
Eliminating this factor we obtain
\begin{align*}
&\mu_{x_{\m+\e_{a+1}}}\left( \mu_{x_{\m+\e_1}}\left( X_{0,\m} \right)\right)\\
&=\left(\prod_{i=a+2}^{a+b}y_{\m+\e_1+\e_i}^{k_i}y_{\m+\e_1-\e_i}^{l_i} \right)^{k_1}y_{\m+\e_1-\e_{a+1}}^{l_{a+1}k_1}
z_{\m+\e_{a+1}}^{k_{a+1}}x_{\m-\e_1}^{l_1}\prod_{i=2}^{a}x_{\m+\e_i}^{k_i}x_{\m-\e_i}^{l_i}\\
&\quad +\left(\prod_{i=2}^{a}y_{\m+\e_{a+1}+\e_i}^{k_i}y_{\m+\e_{a+1}-\e_i}^{l_i} \right)^{k_{a+1}} y_{\m+\e_{a+1}-\e_{1}}^{l_{1}k_{a+1}}z_{\m+\e_1}^{k_1}x_{\m-\e_{a+1}}^{l_{a+1}}\prod_{i=a+2}^{a+b}x_{\m+\e_i}^{k_i}x_{\m-\e_i}^{l_i}. 
\end{align*}
From the above arguments and examples, we can expect the results of mutations.
To give precise statements,  we fix a vertex $\m \in \Z^{a+b}$ and prepare several notations.
Let the subsets of the vertices $U,\, V \subset \Z^{a+b}$ be
\[
U:=\{\m+\e_i,\,\m-\e_i\}_{i=1}^a,\ 
V:=\{\m+\e_i,\,\m-\e_i\}_{i=a+1}^{a+b},
\]
and we define the mapping $\hat{N}:\, \Z^{a+b} \rightarrow \Z_{\ge 0}$ as $\hat{N}(\m+\e_i)=k_i,$\ $\hat{N}(\m-\e_i)=l_i$ and $\hat{N}(\n)=0$ ($\n \ne \m \pm \e_i$) for $i=1,2,...,a+b$ .
We abbreviate $\mu_{x_\n}(\,\cdot\,)$ to $\mu_\n(\,\cdot\,)$ and denote 
$(\mu_\n \circ \mu_{\n'})(\,\cdot\,):=\mu_\n(\mu_{\n'}(\,\cdot\,))$.
For a sequence
\[
\hat{K}:=(\n_1,\n_2,\ldots,\n_f),
\]
we define
\[
\bmu_{\hat{K}}(\,\cdot\,):=(\mu_{\n_f}\circ\cdots\circ\mu_{\n_2}\circ\mu_{\n_1})(\,\cdot\,).
\]
The following lemma is the key to the proof of our main result in this section.
\begin{Lemma}\label{lem_main}
For a sequence $\hat{K}:=(\n_1,\n_2,\ldots,\n_f)$ and the set of its elements $K:=\{\n_1,\n_2,...,\n_f \} (\subset \Z^{a+b})$,
we define the subsets of vertices as
\[\\
A:=U\cap K,\ B:=V\cap K,\ \bar{A}:=U\setminus A,\ \bar{B}:=V\setminus B.
\]
Then it holds that
\begin{equation}\label{propeq1}
\bmu_{\hat{K}}(X_{0,\m})
=C_K(\y) \prod_{\n \in B}z_\n^{\hat{N}(\n)}\prod_{\n \in \bar{A}}x_\n^{\hat{N}(\n)}  +  D_K(\y) \prod_{\n \in A}z_\n^{\hat{N}(\n)}\prod_{\n \in \bar{B}}x_\n^{\hat{N}(\n)},
\end{equation}
where
\begin{equation}
C_K(\y):=\prod_{\n \in A,\,\n' \in \bar{B}}y_{\n+\n'-\m}^{\hat{N}(\n)\hat{N}(\n')},\qquad
D_K(\y):=\prod_{\n \in \bar{A},\,\n' \in B}y_{\n+\n'-\m}^{\hat{N}(\n)\hat{N}(\n')},
\end{equation}
and $C_K(\y)=1$ for $A= \emptyset$ or $\bar{B}= \emptyset$, $D_K(\y)=1$ for $\bar{A}= \emptyset$ or $B= \emptyset$. 
In particular, $\bmu_{\hat{K}}(X_{0,\m})$ does not depend on the order of $\hat{K}$.
\end{Lemma}
\Proof
We prove by induction on the value $f:=|K|$.
Clearly \eqref{propeq1} is true for $f=0$.
(Actually, it follows from the above arguments in this section that \eqref{propeq1} holds for $f=0,1,2$.)

Suppose that \eqref{propeq1} is true for $f \le k$ ($k \ge 0$).
Let 
\[
\hat{K}:=(\n_1,\n_2,\cdots ,\n_k),\quad \hat{L}=(\n_1,\n_2,\cdots ,\n_k,\n_{k+1}).
\]
If $\n_{k+1} \notin U\sqcup V$, then $\bmu_{\hat{L}}(X_{0,\m})=\bmu_{\hat{K}}(X_{0,\m})$ and \eqref{propeq1} is clear.
Therefore, we only consider the case $\n_{k+1} \in U\sqcup V$.
We can assume without loss of generality that $\n_{k+1} \in U$ because of the symmetry of the equation．
Note that $\n_{k+1} \in \bar{A}$.
Since 
\[
\bmu_{\hat{L}}(X_{0,\m})=\mu_{\n_{k+1}}\left(\bmu_{\hat{K}}(X_{0,\m})\right),
\]
\[
\frac{X_{1,\n_{k+1}}|_{y_\m= 0}}{z_{\n_{k+1}}}=\frac{\prod_{i=a+1}^{a+b}y_{\n_{k+1}+\e_i}^{k_i}y_{\n_{k+1}-\e_i}^{l_i}}{z_{\n_{k+1}}},
\]
we have 
\begin{align*}
G&:=\bmu_{\hat{K}}(X_{0,\m})\Big|_{x_{\n_{k+1}}\leftarrow \frac{X_{1,\n_{k+1}}|_{y_\m= 0}}{z_{\n_{k+1}}}   } \\
&=C_K(\y) \left(\prod_{\n \in B}z_\n^{\hat{N}(\n)}\prod_{\n \in \bar{A}\setminus \{\n_{k+1}\}}x_\n^{\hat{N}(\n)}\right) \left(\frac{\prod_{i=a+1}^{a+b}y_{\n_{k+1}+\e_i}^{k_i}y_{\n_{k+1}-\e_i}^{l_i}}{z_{\n_{k+1}}} \right)^{ \hat{N}(\n_{k+1})} \\
&\qquad  \qquad +  D_K(\y) \prod_{\n \in A}z_\n^{\hat{N}(\n)}\prod_{\n \in \bar{B}}x_\n^{\hat{N}(\n)}\\
&=\frac{1}{z_{\n_{k+1}}^{\hat{N}(\n_{k+1})}}\left\{C_K(\y)\left(\prod_{\n \in B}z_\n^{\hat{N}(\n)}\prod_{\n \in \bar{A}\setminus \{\n_{k+1}\}}x_\n^{\hat{N}(\n)}\right) \left(\prod_{i=a+1}^{a+b}y_{\n_{k+1}+\e_i}^{k_i}y_{\n_{k+1}-\e_i}^{l_i} \right)^{ \hat{N}(\n_{k+1})} \right.\\
&\qquad \qquad \left. +  D_K(\y) \prod_{\n \in A\sqcup \{\n_{k+1}\}}z_\n^{\hat{N}(\n)}\prod_{\n \in \bar{B}}x_\n^{\hat{N}(\n)}\right\},\\
\end{align*}
\begin{align*}
\left(\prod_{i=a+1}^{a+b}y_{\n_{k+1}+\e_i}^{k_i}y_{\n_{k+1}-\e_i}^{l_i} \right)^{ \hat{N}(\n_{k+1})} 
&=\prod_{\n \in V}y_{\n+\n_{k+1}-\m}^{\hat{N}(\n)\hat{N}(\n_{k+1})     }\\
&=\left(\prod_{\n \in B}y_{\n+\n_{k+1}-\m}^{\hat{N}(\n)\hat{N}(\n_{k+1})     }    \right)  \left(\prod_{\n \in \bar{B}}y_{\n+\n_{k+1}-\m}^{\hat{N}(\n)\hat{N}(\n_{k+1})     }    \right)  
\end{align*}
and 
\[
D_K(\y)=\prod_{\n \in \bar{A},\,\n' \in B}y_{\n+\n'-\m}^{\hat{N}(\n)\hat{N}(\n')}
=\left(\prod_{\n \in \bar{A}\setminus\{\n_{k+1}\},\,\n' \in B}y_{\n+\n'-\m}^{\hat{N}(\n)\hat{N}(\n')} \right)
\left(\prod_{\n \in B}y_{\n+\n_{k+1}-\m}^{\hat{N}(\n)\hat{N}(\n_{k+1})     }    \right). 
\]
Removing the common factor $\left(\prod_{\n \in B}y_{\n+\n_{k+1}-\m}^{\hat{N}(\n)\hat{N}(\n_{k+1})     }    \right) $,
we find that
\begin{align*}
\bmu_{\hat{L}}(X_{0,\m})&=C_L(\y)\prod_{\n \in B}z_\n^{\hat{N}(\n)}\prod_{\n \in \bar{A}\setminus \{\n_{k+1}\}}x_\n^{\hat{N}(\n)}+D_L(\y)\prod_{\n \in A\sqcup \{\n_{k+1}\}}z_\n^{\hat{N}(\n)}\prod_{\n \in \bar{B}}x_\n^{\hat{N}(\n)},\\
C_L(\y)&=\left(\prod_{\n \in \bar{B}}y_{\n+\n_{k+1}-\m}^{\hat{N}(\n)\hat{N}(\n_{k+1})     }    \right)  C_K(\y)=\prod_{\n \in A\sqcup \{\n_{k+1}\},\,\n' \in \bar{B}}y_{\n+\n'-\m}^{\hat{N}(\n)\hat{N}(\n')},\\
D_L(\y)&=\prod_{\n \in \bar{A}\setminus\{\n_{k+1}\},\,\n' \in B}y_{\n+\n'-\m}^{\hat{N}(\n)\hat{N}(\n')}.
\end{align*}
Hence, \eqref{propeq1} is true for $f=k+1$.
By the induction hypothesis, \eqref{propeq1} is true for arbitrary $\hat{K}$.
\qed

Let us denote by $\bmu_{\infty}$ the mutation of all the vertices.
Since
$A = U$, $B = V$, $C_K = 1$, $D_K = 1$ for the mutation $\bmu_{\infty}$,
we obtain our main theorem in this section:
\begin{Theorem}
The extended discrete Toda lattice equation \eqref{pNDToda_polinear_eq} has the LP algebraic structure:
\[
\bmu_{\infty}\left(X_{0,\m} \right)=X_{2,\m}.
\]
\end{Theorem}

%
%


\section{Conclusion}
In this paper, we introduced a higher dimensional extension to the coprimeness-preserving two-dimensional discrete Toda lattice equation \eqref{pNDToda_polinear_eq}.
The equation \eqref{pNDToda_polinear_eq} has the Laurent property, the irreducibility and the coprimeness. The equation is non-integrable in terms of exponential degree growth,
nevertheless, it has the coprimeness property just like most of the discrete integrable systems.
In terms of integrability, one of the highest dimensional lattice equations is the
two-dimensional discrete Toda lattice equation, which is defined over $\mathbb{Z}^3$. However, if the integrability condition is weakened to the coprimeness-preserving condition, we can construct an equation over the lattice of the dimension higher than three such as \eqref{pNDToda_polinear_eq}.
In the last section, we proved that \eqref{pNDToda_polinear_eq} can be expressed as a mutation of the Laurent phenomenon algebra. The Laurent phenomenon algebra is, roughly speaking,  an extension of the theory of cluster algebras  and is expected to generate
wider class of discrete equations including non-integrable ones.
It is an interesting task to investigate the co-relations between the Laurent phenomenon algebra and the coprimeness property.
When we study the reductions of the equation \eqref{pNDToda_polinear_eq}, it yields
various coprimeness-preserving equations over lower dimensional lattices.
One of the future works is to give the full answer to the problem relating coprimeness-preserving equations to integrability.
To achieve this goal we wish to classify the coprimeness-preserving equations into some hierarchy of reduction.
The nature of such coprimeness-preserving equations is not well-understood compared to the integrable ones, and we wish to further study equations with the coprimeness property but without integrability.

\section*{Acknowledgement}
The authors are grateful to Prof. R. Willox for useful comments.
The present work is partially supported by KAKENHI Grant Numbers 16H06711 and 17K14211. 

\appendix
\section{Factorization lemma}
Let us reproduce a lemma on how a Laurent polynomial is factorized
under a change of variables, when the variable transformation is fairly simple.
The situation we often encounter is that, the conditions on the following lemma \ref{lem2previous} are satisfied thanks to the Laurent property and the invertibility of the equation that we investigate.
\begin{Lemma}[\cite{dKdVSC2}]
\label{lem2previous}
Let $M$ be a positive integer and let $\{p_1,p_2,\cdots,p_M\}$ and $\{q_1,q_2,\cdots ,q_M\}$ be two sets of independent variables with the following properties:
\begin{align*}
p_j &\in \mathbb{Z}\left[ q_1^{\pm}, q_2^{\pm},\cdots ,q_M^{\pm}\right], 
q_j \in \mathbb{Z}\left[ p_1^{\pm}, p_2^{\pm},\cdots ,p_M^{\pm}\right], \\
q_j&\ \mbox{is irreducible as an element of}\ \mathbb{Z}\left[ p_1^{\pm}, p_2^{\pm},\cdots ,p_M^{\pm}\right],  \notag
\end{align*}
for $j=1,2,\cdots, M$.
Let us take an irreducible Laurent polynomial
\[
f(p_1,\cdots, p_M)\in \mathbb{Z}\left[ p_1^{\pm}, p_2^{\pm},\cdots ,p_M^{\pm}\right],
\]
and another (not necessarily irreducible) Laurent polynomial
\[
g(q_1,\cdots, q_M) \in \mathbb{Z}\left[ q_1^{\pm}, q_2^{\pm},\cdots ,q_M^{\pm}\right],
\]
which satisfies $f(p_1,\cdots,p_M)=g(q_1,\cdots, q_M)$.
In these settings, the function $g$ is decomposed as
\[
g(q_1,\cdots, q_M)=p_1^{r_1}p_2^{r_2}\cdots p_M^{r_M}\cdot \tilde{g}(q_1,\cdots, q_M),
\]
where $r_1,r_2, \cdots, r_M\in\mathbb{Z}$ and $\tilde{g}(q_1,\cdots,q_M)$ is irreducible in $\mathbb{Z} \left[ q_1^{\pm}, q_2^{\pm},\cdots ,q_M^{\pm}\right]$.
\end{Lemma}
The underlying idea is the fact in algebra that the localization of a unique factorization domain preserves the irreducibility of its elements.
Proof is found in reference \cite{dKdVSC2}.

\section{Coprimeness of two-dimensional Toda lattice} \label{2dfactorappendix}
Let us briefly review the results on the two-dimensional Toda lattice and its extension \eqref{pDToda_polinear_eq}:
\begin{equation*}
\tau_{t+1,n,m+1}\tau_{t-1,n+1,m}=\tau_{t,n+1,m}^{k_1}\tau_{t,n,m+1}^{k_2}+\tau_{t,n,m}^{l_1}\tau_{t,n+1,m+1}^{l_2}\qquad (k_i, l_i \in \Z_+).
\end{equation*}
It is proved that \eqref{pDToda_polinear_eq} has the Laurent, the irreducibility and the coprimeness properties, if the right hand side of the equation is not factorizable:
\begin{Theorem}[\cite{2dtoda1}] \label{2dthm}
Let us assume that the greatest common divisor of $(k_1,k_2,l_1,l_2)$ is a non-negative power of $2$. Then each iterate $\tau_{t,\n}$ of equation \eqref{pDToda_polinear_eq} is an irreducible Laurent polynomial of the initial variables
$\left\{\tau_{0, n,m},\,\tau_{1, n,m}\, |\, n,m\in\mathbb{Z}\right\}$.
Moreover, every pair of the iterates is always co-prime.
\end{Theorem}
Note that the condition $\operatorname{GCD}(k_1,k_2,l_1,l_2)= 2^k$ with $k\ge 0$ is equivalent to
the irreducibility of $P^{k_1}Q^{k_2}+R^{l_1}S^{l_2}$ in $\mathbb{Z}[P,Q,R,S]$.
Now a slightly more general result is proved:
\begin{Theorem}[\cite{2dTodaFactorize}] \label{factorizethm}
Each iterate $\tau_{t,\n}$ of equation \eqref{pDToda_polinear_eq} is a Laurent polynomial of the initial variables.
Moreover, every pair of the iterates is always co-prime.
\end{Theorem}
Here the condition on the previous theorem \ref{2dthm} is omitted. Therefore the irreducibility is lost in theorem \ref{factorizethm}. Nevertheless the Laurent property and the coprimeness are still satisfied.


\section{Basic facts on LP algebras}\label{LP_Basic_facts}
Let us briefly review the notion of the Laurent Phenomenon algebra based on \cite{LP} by Lam and Pylyavskyy.
Laurent Phenomenon algebras have been proposed as one type of generalizations to cluster algebras.

For simplicity let us always study polynomials and Laurent polynomials over $\Z$.
We call a pair $\mF:=(\x,\F)$ of variables $\x$ and polynomials $\F$  as a seed. Here
\[
\x=(x_1,x_2,...,x_N),\quad \F=(F_1,F_2,...,F_N),
\]
where $F_i \in \Z[\x]$. Each variable $x_i$ is called a cluster variable and each polynomial $F_i$ is called an exchange polynomial.
They are supposed to have the following two properties:
\begin{itemize}
\item[(LP1)] Each $F_i \in \Z[\x]$ is irreducible and is not divisible by any $x_j$ $(j=1,2,\cdots, N)$.
\item[(LP2)] Each $F_i$ is independent of $x_i$.
\end{itemize}
The exchange Laurent polynomials $\hat{\F}=(\hat{F}_1,\hat{F}_2,...,\hat{F}_N)$ $(F_i \in \Z[\x^\pm])$ are defined to satisfy the following two properties:
\begin{itemize}
\item[(LP3)]  $\hat{F}_j=u_j(\x)^{-1}F_j$, where $u_j(\x) \in \Z[\x]$ is a monic monomial independent of $x_j$.
\item[(LP4)]  $\DIS \hat{F}_i\big|_{x_j \leftarrow F_j/x_j'}$ belongs to $\Z[x_1^\pm,...,x_{j-1}^\pm,{x_j'}^\pm,x_{j+1}^\pm,...,x_N^\pm]$ and is not divisible by $F_j$ in the Laurent polynomial ring.
\end{itemize}
Let us give one simple example which is quoted from \cite{LP}.
\begin{Example}\label{example1}
Let us consider the initial seed $\x=(a,b,c)$, 
\[
F_a=b+1,\ F_b=(a+1)^2+c^2,\ F_c=b^2+b+a^3+a^2.
\]
Since
\begin{align*}
F_a\big|_{b\leftarrow \frac{F_b}{b'}}&=\frac{(a+1)^2+c^2}{b'}+1,\\
F_b\big|_{a \leftarrow \frac{F_a}{a'}}&=\left\{\frac{b+1}{a'} +1\right\}^2+c^2,\\
F_b\big|_{c \leftarrow \frac{F_c}{c'}}&=(a+1)^2+\left\{\frac{b^2+b+a^3+a^2}{c'}\right\}^2\\
F_c\big|_{a \leftarrow \frac{F_a}{a'}}&=b^2+b+\left(\frac{b+1}{a'} \right)^3+\left( \frac{b+1}{a'}\right)^2 \\
&=(b+1)\left[ b + \frac{(b+1)^2}{{a'}^3}+\frac{b+1}{{a'}^2}\right],\\
F_c\big|_{b\leftarrow \frac{F_b}{b'}}&=\left\{ \frac{(a+1)^2+c^2}{b'} \right\}^2+\frac{(a+1)^2+c^2}{b'} +a^3+a^2,
\end{align*}
we have $\hat{F}_a=F_a$, $\hat{F}_b=F_b$ and $\hat{F}_c=a^{-1}F_c$．
\end{Example}
The well-definedness of these definitions are assured by the lemma \ref{lem2}.
\begin{Lemma}\label{lem2}
The sets $\{F_1,F_2,...,F_N\}$ and $\{\hat{F}_1,\hat{F}_2,...,\hat{F}_N\}$ define each other bijectively.
\end{Lemma}
Next let us define an operation called a mutation to a seed.
\begin{Definition}
For a fixed $i \in\{1,2,...,N\}=:[N]$, let us define the mutation $(\x',\F')=\mu_i(\x,\F)$ of the seed $(\x,\F)$ as follows:
New cluster variables are
\begin{align*}
x_i'&=\hat{F}_i/x_i, \\
x_j'&=x_j \quad (j \ne i).
\end{align*}
New exchange Laurent polynomials are defined as follows:
first $F_i'(\x')=F_i(\x)$.
For $j \ne i$, $F_j'$ is defined as
\begin{itemize}
\item If $F_j$ is independent of $x_i$, $F_j'(\x')=F_j(\x)$.
\item If $F_j$ depends on $x_i$, we define $G_j$ as:
\[
G_j:=F_j \Big|_{x_i \leftarrow \frac{\hat{F}_i|_{x_j \leftarrow 0}}{x_i'}}.
\]
$G_j$ is well-defined because $\hat{F}_i$ cannot have negative power of $x_j$ when $F_j$ depends on $x_i$ (which is not trivial but the proof is omitted here) and thus $\DIS \frac{\hat{F}_i\big|_{x_j \leftarrow 0}}{x_i'} $ is not divergent.
\item $F_j'$ is constructed from $G_j$ as follows: let us use the notation $\x_i=\x\setminus \{x_i\}$, $\x_{ij}:=\x\setminus \{x_i,x_j\}$ and  define the polynomial $H_j$ by removing all the common factors of $G_j$ and $\hat{F}_i\big|_{x_j \leftarrow 0} \in \Z[\x_{ij}]$ from $G_j$.
Since $H_j \in \Z[{\x'}^\pm]$, there exists a monomial $M \in \Z[\x']$ such that $M G_i$ satisfies (LP2). For this $M$ we define $F_j'(\x'):=MG_j$.
\end{itemize}
The following propositions hold for the mutations.
\end{Definition}
\begin{Example}\label{example2}
In Example \ref{example1}, let us apply the mutation at $c$ to the initlal seed.
Let $c'=d$.
From $d=\displaystyle \frac{\hat{F}_c}{c}$ we have $d=\displaystyle \frac{b^2+b+a^3+a^2}{ac}$.
Since $F_a$ is independent of $c$, we have $F_a'=F_a$. From
\begin{align*}
G_b&=F_b\big|_{c\leftarrow \frac{F_c |_{b=0}}{d}}=(a+1)^2+\left(\frac{a^3+a^2}{ad}\right)^2\\
&=(a+1)^2\left\{1+\left(\frac{a}{d}\right)^2\right\}=\frac{(a+1)^2}{d^2}(a^2+d^2),
\end{align*}
$(a+1)^2$ is a common factor of $F_c\big|_{0\leftarrow b}$ and $G_b$. Removing the denominator we have $F_b'=a^2+d^2$.
Therefore the new seed is
\[
\left\{ (a, b+1), \, (b,a^2+d^2), \, (d, b^2+b+a^3+a^2) \right\}.
\]
\end{Example}
\begin{Proposition}\label{prop1}
The new seed $(\x',\F')=\mu_i(\x,\F)$ satisfies (LP1) and (LP2).
\end{Proposition}
We write $(\x,\F)\longrightarrow (\x',\F')$ to denote the mutation.
\begin{Example}\label{example3}
We apply the mutation at $d$ for Example \ref{example2}. Since
\[
F_d\big|_{a \leftarrow \frac{F_a}{x}}=b^2+b+\left(\frac{b+1}{x}\right)^3+\left(\frac{b+1}{x}\right)^2,
\]
we have $\hat{F}_d=a^{-1}F_d$ and
\[
d'=\frac{\hat{F}_d}{d}=\frac{b^2+b+a^3+a^2}{ad}=c.
\]
From $\hat{F}_d\big|_{b=0}=a^2+a$,
\[
G_b'=F_b'\big|_{d\leftarrow \frac{a^2+a}{c}}=a^2+\left(\frac{a^2+a}{c}\right)^2
=\frac{a^2}{c^2}\left\{(a+1)^2+c^2
\right\}.
\]
Thus $F_b''=(a+1)^2+c^2=F_b$. Since $F_a''=F_a'=F_a$, we have
\[
\mu_d\left(\mu_c(\x,\F)\right)=(\x,\F),
\]
which directly shows Proposition \ref{prop2} for this particular example.
It is worth noting that the mutations are generally non-commutative:
for example, if we set $f:=b''$ we have
\[
\mu_b(\mu_c(F_a))=f+d^2,\quad \mu_c(\mu_b(F_a))=1+f+d^2.
\]
In our manuscript, the LP algebra corresponding to the Toda type equation is shown to be well-defined independent of the order of mutations.
\end{Example}
\begin{Proposition}\label{prop2}
If $(\x',\F')=\mu_i(\x,\F)$, we have $(\x,\F)=\mu_i(\x',\F')$.
\end{Proposition}
Finally let us define the LP algebra constructed from a seed $(\x,\F)$.
\begin{Definition}
Let us fix a seed $(\x,\F)$, which shall be called the initial seed and let $X((\x,\F))$ be the set of cluster variables obtained by applying successive mutations to the initial seed. Then the LP algebra associated with the initial seed is
\[
\mathcal{A}((\x,\F)):=\mathbb{Q}[x\, |\, x\in X((\x,\F))].
\]
\end{Definition}
The following theorem on the Laurent property forms the basis of application of LP algebras to discrete integrable systems.
\begin{Theorem}\label{theorem1}
Every cluster variable $\y\in X((\x,\F))$ belongs to $\mathbb{Z}[\x^{\pm}]$.
\end{Theorem}
%
%
%

\end{document}